\documentclass[hackGuillaume]{svjour3} % hackGuillaume natbib
\usepackage[utf8]{inputenc}
\usepackage[T1]{fontenc}
\usepackage[english]{babel}
\usepackage{times,mathptmx}

\usepackage{graphicx}
\usepackage[usenames,svgnames,table]{xcolor}
\usepackage[colorlinks,allcolors=MediumBlue,breaklinks,hyperfootnotes=true,
	pdftitle={Decontamination of the scientific literature},
	pdfauthor={Guillaume Cabanac},
	pdfsubject={Application to the 2022 IUF Chair Programme},
	pdfkeywords={research integrity, scientific literature, whistleblowing, errors, fraud}
]{hyperref}

\usepackage[natbibapa]{apacite}    % instead of passing natbib to svjour3
\newcommand{\doi}[1]{\href{https://doi.org/#1}{#1}}
\usepackage{underscore}
\emergencystretch=\maxdimen
\begin{document}
\journalname{Application to the 2022 \href{https://en.wikipedia.org/wiki/Institut_Universitaire_de_France}{IUF Chair} Programme}

\title{Decontamination of the scientific literature}
%\subtitle{}

\author{Guillaume Cabanac}
\authorrunning{G. Cabanac}

\institute{
G. Cabanac\at
University of Toulouse, Computer Science Department, IRIT UMR 5505 CNRS, 31062 Toulouse, France\\
Institut Universitaire de France (IUF), 75005 Paris, France\\
\email{guillaume.cabanac@univ-tlse3.fr}\\
ORCID: \href{http://orcid.org/0000-0003-3060-6241}{0000-0003-3060-6241}
}

% Started writing on 27-SEP-2021
\date{\href{https://web.archive.org/web/2021/https://www.iufrance.fr/campagne-de-selection-des-nouveaux-membres-iuf.html}{Submitted}: November 2, 2021 $\cdot$ \href{https://www.iufrance.fr/les-membres-de-liuf/membre/2427-guillaume-cabanac.html}{Granted}: May 2, 2022 $\cdot$ Preprinted: October 28, 2022}
\maketitle 

\urlstyle{same}

\begin{abstract}
    Research misconduct and frauds pollute the scientific literature.
	Honest errors and malevolent data fabrication, image manipulation, journal hijacking, and plagiarism passed peer review unnoticed.
	Problematic papers deceive readers, authors citing them, and AI-powered literature-based discovery.
	Flagship publishers accepted hundreds flawed papers despite claiming to enforce peer review.
	This application ambitions to decontaminate the scientific literature using curative and preventive actions.
\end{abstract}


%=========================================================================================================================
\section{Context: Fraud and misconduct undermine trust in science}

In 2020 the scientific community reached a new world record despite the global COVID19 pandemic.
Scholarly knowledge production peeked at \href{https://app.dimensions.ai/discover/publication?or_facet_publication_type=article&or_facet_publication_type=proceeding&or_facet_year=2020}{5 million peer-reviewed articles} as per the Dimensions bibliographic database \citep{HerzogEtAl2020}.
Year after year, research production grows in terms of publication counts \citep[p.~36]{SoeteEtAl2015} and geographic footprints \citep{MaisonobeEtAl2018a}.
The `publish or perish' atmosphere, which is already decades-old \citep{Cabanac2018a}, intensifies when scientific papers once viewed as \emph{knowledge} units are increasingly considered as \emph{accounting} units \citep{Gingras2020}.
Some institutions---countries, even---have implemented financial incentives to reward authors who manage to publish in the most highly-regarded journals \citep{Chen2019,Lin2013}.
Such incentives contravene the `disinterestedness' norm of the Ethos of Science \citep{Merton1942} and they indirectly fostered fraud and misconduct to `game the metrics' \citep{BiagioliAndLippman2020} with documented cases of:
\begin{itemize}
	\item `Paper mills' as industrialised ghostwriting where companies churn pseudo-scientific papers and sell authorship positions \citep{ElseAndVanNoorden2021,Oransky2021b}.

	\item Publication with predatory publishers implementing weak peer-review \citep{GrudniewiczEtAl2019,Faulkes2021}.

	\item Hijacked journals fooling authors by replicating logos and names of legitimate journals \citep{Abalkina2021}.

	\item Identity theft to impress editors/reviewers with prestigious affiliations \citep{Barbash2014} and fool editorial assessment with forged researcher identifiers \citep{TeixeiraDaSilva2021}.

	\item Data fabrication as in the Surgisphere scandal during the pandemic \citep{Piller2020}.
	
	\item Image manipulation, especially in the health sciences \citep{Shen2020}.
	
	\item Undeclared conflicts of interests casting doubts on authors' motivations and their results \citep{LewisonAndSullivan2015,ConfrontingCOI2018}.
	
	\item Computer-generation of nonsensical text to pad papers or even generate entire studies \citep{CabanacAndLabbe2021a}.

	\item Plagiarism of published articles \citep{Pupovac2021} using thesaurus-based paraphrasing \citep{CabanacEtAl2021c}.
\end{itemize}

It is of critical importance to find and remove such pollution from the literature, as erroneous publications potentially deceive:
\begin{itemize}
	\item \emph{Readers} whose expertise in the field might not be high enough to detect errors.  Readers turned authors may even propagate the errors when citing flawed papers.
	
	\item \emph{Meta-researchers} who perform systematic reviews of the literature that may contain flawed papers.  What if the evidence used to recommend a medical treatment is based on an inaccurate study?  Preprints proved challenging for living systematic reviews \citep{OikonomidiEtAl2020} and so are erroneous publications.
	
	\item \emph{AI-powered software} performing literature-based discovery \citep{BruzaAndWeeber2008} and writing literature reviews automatically \citep{BetaWriter2019}. 
\end{itemize}

While retractions of problematic publications are on the rise \citep{BrainardAndYou2018,AbritisEtAl2020,Sharma2021} the public is entitled to ask: How many problematic papers go unnoticed?
This sadly contributes to the `reproducibility crisis' in various research areas \citep{Baker2016}, which affects the trust people place in science.

In October 2021, the Institut Universitaire de France (IUF) organised its annual congress titled \emph{Post-truth? Credibility of scientific research in a time of ``alternative facts''} to foster research on this concerning issue.
The talks I attended laid a dark landscape of science in our post-truth era; they convinced me to apply to the IUF Chair programme.

%=========================================================================================================================
\section{This IUF application: Decontamination of the scientific literature}
	During my 5-year tenure as Junior IUF member, I plan to tackle the issue of pollution staining the scientific literature, a current concern shared with the \emph{Office Français de l'Intégrité Scientifique} \citep{OFIS2021}.
	My plan is twofold.
	Its \emph{curative} part deals with the problematic papers that are already published: we need to find them, report them, and get them down.
	Its \emph{preventive} part ambitions to anticipate the new forms of misconduct and react as soon as spoiled literature pops up.

%~~~~~~~~~~~~~~~~~~~~~~~~~~~~~~~~~~~~~~~~~~~~~~~~~~~~~~~~~~~~~~~~~~~~~~~~~~~~~~~~~~~~~~~~~
\subsection{Curative approach for decontaminating the scientific literature}
	The pollution staining the scientific literature is here to stay until responsible scientists or publishers spot it and take action.
	Individually or as team member, I have initiated three concrete actions to wipe the literature from erroneous studies.
	The IUF Chair would contribute to upscale and enhance these ongoing endeavours that already yielded substantial results.

%----------------------------------------------------------------------------------
\subsubsection{Detection and correction of the pollution by paper mills in oncology}
	In 2014, a Professor of molecular oncology at the University of Sydney noticed errors in the genetic sequences reported in articles published by a variety of journals.
	These papers would claim that a given DNA sequence 1)~targets human gene~X or 2)~targets none of the human genes.
	These claims form the basis of gene knockdown experiments and any inaccuracy invalidates any results presented.
	Meticulous readers would check these claims with the \href{https://blast.ncbi.nlm.nih.gov}{BLAST} search tool hosted by NCBI.
	This Prof.\ Jennifer Byrne did, only to find hundreds of erroneous claims in \emph{de facto} erroneous papers.
	Her sleuth work gained high recognition in 2017: The \emph{Nature} journal listed her in the Top~10 of people who mattered that year \citep{Philipps2017}.
	
	Prof.\ Byrne teamed up with Dr.\ Cyril Labbé, an Associate Professor from Grenoble to build the \href{http://scigendetection.imag.fr/TPD52/Vb/}{Seek\&Blastn software} that automates the screening of DNA sequences extracted from papers \citep{LabbeEtAl2019}.
	I joined the Byrne--Labbé collaboration and we designed an Information Retrieval task and evaluation benchmark to assess the effectiveness of error detection in papers reporting genetic sequences \citep{LabbeEtAl2020a}.
	We screened thousands of papers and reported the problematic ones to editors-in-chief and publishers.
	Problems relate to unsupported claims (targetting/non targetting DNA sequences) and untraceable source of genetic materials used (e.g., unknown ‘Hollybio’ company named as supplier of biologic material).
	Most suspect papers follow a common template (e.g., study of the effect of gene X on organ/condition Y) that is typical of the `paper mills' output \citep{ElseAndVanNoorden2021}.
	Despite our team's continuing efforts, formal retractions are long to come---when they do come \citep{ByrneEtAl2021a}.
	
	Our latest research revealed the extent of problematic papers in two oncology journals that we screened cover-to-cover thus checking 13,700 nucleotide sequences in 3,400 papers.
	The 21\% error-rate is unacceptably high for this scholarly literature.
	In addition, the positive reception of these erroneous papers (17k citations to 712 papers) suggests citation manipulation at scale \citep{ParkEtAl2021a}.
	This alarming result has been profiled in the \emph{News} section of \emph{Nature} \citep{Else2021a} and we currently analyse the suspect citation networks to hopefully uncover any citation cartels.

%----------------------------------------------------------------------------------
\subsubsection{Detection and reporting of hijacked and hacked scientific journals}
	Predatory publishers are known to sell quick time-to-market with little to no peer review \citep{GrudniewiczEtAl2019}.
	The reputable titles and logos of established journals get hijacked when scammers copy them to create fake lookalikes \citep{Abalkina2021}.
	These fool the inattentive---naive---authors who believe they submitted to a reputable venue.
	Crooked authors also submit weak or even nonsensical computer-generated papers knowingly.

	Flagship publishers were considered immune to nonsensical submissions.
	It was especially true for `elite' journals with Impact Factors---20\% only have one.
	And yet, we found that an Elsevier journal with Impact Factor, \emph{Microprocessors and Microsystems}, had been compromised in 2021.
	It had published no less than 400 problematic articles when we released our whistleblowing study reporting tortured phrases and shrinking durations of editorial assessments \citep{CabanacEtAl2021c}.
	Elsevier acknowledged the errors and asked experts to reassess all these suspect articles \citep{Marcus2021a,Marcus2021b,Else2021b}.

	We also contributed to uncover a similar integrity breach at Springer's \emph{Arabian Journal of Geosciences}.
	This journal with Impact Factor had published more than 400 absurd papers when we reported this case to the public \citep{Oransky2021a,Oransky2021c}.
	Analysing the timelines of editorial assessments I highlighted an unexpectedly steep time-to-market shrinkage starting in 2021.\footnote{\url{https://github.com/gcabanac/editorial-assessment}}
	Responsible publishers should monitor and investigate such sudden changes in their production to prevent other nonsensical `bubbles' bursting after hundreds problematic papers were published.

%----------------------------------------------------------------------------------
\subsubsection{Detection and reporting of nonsensical algorithmically generated articles}
	Highly-regarded publishers in engineering, such as ACM and IEEE, have published and sold meaningless computer-generated papers.
	We designed an algorithm to comb the literature for fraudulent papers generated with probabilistic context-free grammars such as \href{https://pdos.csail.mit.edu/archive/scigen/}{SCIgen} and \href{https://thatsmathematics.com/mathgen/}{Mathgen}.
	We found 262 such papers, 197 of these had not been retracted despite being published for many years \citep{CabanacAndLabbe2021a}.
	The screening process runs daily and results appear at the \emph{Problematic Paper Screener}\footnote{\url{https://www.irit.fr/~Guillaume.Cabanac/problematic-paper-screener}} (\href{https://www.irit.fr/~Guillaume.Cabanac/problematic-paper-screener}{PPS}) a public website I have been developing.
	This work was profiled in the \emph{News} section of \emph{Nature} \citep{VanNoorden2021a}.
	As advised by \citet{OFIS2021}, I authored 200 post peer-review comments publicly available on the \href{https://www.pubpeer.com}{PubPeer} platform \citep{BarbourAndStell2020}.
	Some comments have led publishers to retract their papers; crooked authors occasionally posted vulgar replies without providing any acceptable rationale for their wrongdoings \citep{Marcus2021c}.

	My main contribution to error detection has been to collaboratively coin the `tortured phrases' concept, find thousands of such `tortured' papers, and issue an Open Call for Investigation to the scientific community \citep{CabanacEtAl2021c}.
	The new form of plagiarism we identified involves paraphrasing to evade plagiarism detection.
	Fraudsters copy texts from various sources, paraphrase them, and paste them to their own articles.
	Paraphrasing entails the use of a thesaurus to change original words into synonyms.
	Established phrases such as `artificial intelligence' and `Naïve Bayes' (a machine learning technique named after Reverend Thomas Bayes) get synonymised as chimeric `counterfeit consciousness' and `innocent/credulous Bayes.'
	Crooked authors neither notice nor bother to correct these abnormalities that get published by flagship publishers.
	We view `tortured phrases' as tips to focus our attention on suspect papers \citep{CabanacEtAl2022a}.
	The \emph{News} section of \emph{Nature} profiled our efforts to unveil this emerging form of plagiarism \citep{Else2021b}.
	
\begin{figure}[h]\centering
		\setlength{\fboxsep}{0pt}
		\fcolorbox{gray}{white}{\includegraphics[width=\linewidth,trim={0 0 28cm 0},clip]{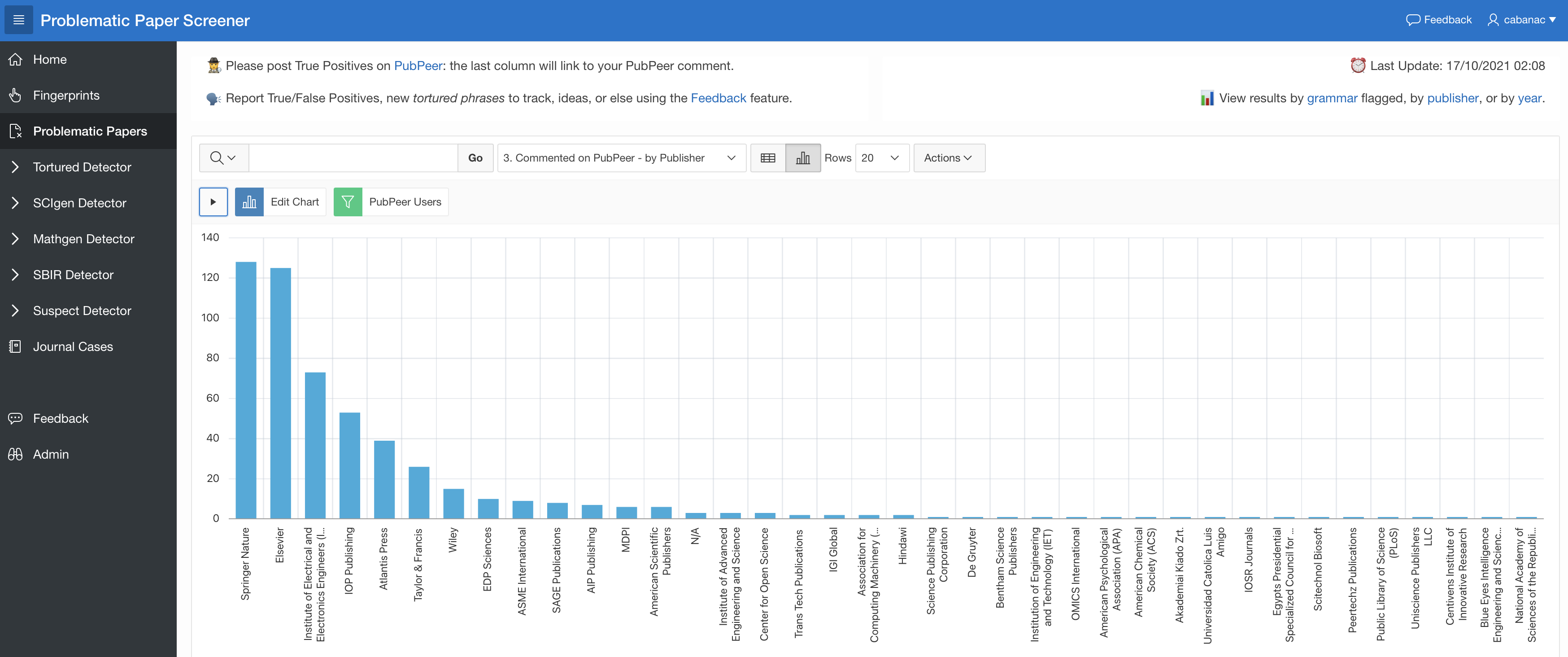}}
		\caption{Per publisher breakdown of the 602 flawed papers flagged by the \href{https://www.irit.fr/~Guillaume.Cabanac/problematic-paper-screener}{PPS} and commented on PubPeer. The `\href{https://dbrech.irit.fr/pls/apex/f?p=9999:5}{fingerprints}' tab lists the tracked tortured phrases (e.g., irregular timberlands) and established wordings that readers expect in the literature (e.g., random forests).}\label{fig:pps}
	\end{figure}

	Our Cabanac--Labbé--Magazinov team detects `tortured' papers by screening the entire scientific literature daily.
	We (re)assess the publications featuring tortured phrases and the \href{https://www.irit.fr/~Guillaume.Cabanac/problematic-paper-screener}{PPS} invites the scientific community at large to likewise.
	Post-publication evaluation reports are crowdsourced from PubPeer and we implement a snowballing approach to integrate newly found `tortured phrases' for forthcoming screening batches.
	Research integrity sleuths and whistleblowers have joined forces, one of the most active being Elisabeth Bik whose detective work was profiled in \citep{Shen2020}.
	As of October 2021, we have flagged 2,225 problematic papers (with 1,823 `tortured papers') 602 of which being commented on PubPeer by 60 individuals (Fig.~\ref{fig:pps}).

	We found evidence that the latest advances in text generation using deep neural networks are being diverted to produce and publish nonsensical articles.
	Running 140k paper abstracts through the GPT-2 detector, we found high concentrations of abstracts exhibiting a GPT nature (i.e., `fake' texts in GPT parlance) in the corpus of Elsevier publications of 2021 \citep[][Tab.~6]{CabanacEtAl2021c}.
	This concerning result calls for an in-depth study to determine how these were produced (machine translation, text generation, paraphrasing or else?) and assess the soundness of the associated publications.

%----------------------------------------------------------------------------------------------------
\subsection{Preventive approach to decontaminate the scientific literature}
	While the \emph{curative} part of my project targets the problematic papers already published, the \emph{preventive} part ambitions to prevent new problematic articles from integrating the scientific record in the first place.
	The surge of GPT-like papers we observed calls for an interdisciplinary sleuth work supported by skills in computer science, natural language processing, and artificial intelligence \citep{VenemaEtAl2020,Hutson2021}.
	Now obsolete text generation models like GPT-2 have been used to generate scientific papers \citep{Lang2019}. 
	One can only fear that the availability of enhanced models \citep[e.g., GPT-J was trained on scientific corpora, see][]{Wiggers2021} will encourage crooked authors even more.

	How to anticipate the next stream of flawed papers?
	I envision to approach computational linguists to work on the automatic detection of unexpected phrases appearing in papers without being common in the associated journal or field.
	In addition, we should take advantage of emerging deceptive ways to publish nonsense being flagged by epistemic activists like Elisabeth Bik and other users of the \href{https://www.irit.fr/~Guillaume.Cabanac/problematic-paper-screener}{PPS}.
	Following \citet{Zuckerman2020}, I plan to keep studying the post peer-review reports and design algorithmic counterattacks to flag misconduct at the global literature scale.
	We should also systematically monitor publicly available data about the publishers' journal-wise output and post-peer review reports.

%=========================================================================================================================
\section{Expected contributions and outcomes}
	During the 5-year tenure as IUF Chair, I wish to achieve the following 5 goals.

%----------------------------------------------------------------------------------------------------
\subsection{Designing of enhanced algorithms to comb the literature for errors}\label{sec:algo}
	We approached the detection of algorithmically-generated papers and tortured papers by screening the literature for certain fingerprint--queries \citep{CabanacAndLabbe2021a}.
	The ongoing crowdsourcing operation supported by our \href{https://www.irit.fr/~Guillaume.Cabanac/problematic-paper-screener}{PPS} website reveals new tortured phrases every day.
	I plan to use text mining to complement this qualitative approach by digging suspect wordings from the corpora under study.
	An exploratory study using pointwise mutual information \citep{Bouma2009} yielded promising results.
	Some phrases are acceptable in some fields while nonsensical in others; one needs to account for intra-field likelihood to reduce the false positive detection rate.

	Some forms of error detection performed manually should benefit from automation.
	We have reported how misidentified cell lines\footnote{See \url{https://iclac.org/databases/cross-contaminations/}.} were misused in selected biomedical publications \citep{ByrneEtAl2021a,ParkEtAl2021a}.
	Biological research using (or citing papers using) misidentified materials are not only a waste of time and resources but also a risk for health.
	The time is ripe to upscale our current digital test-tube to screen the \emph{entire} literature, analyse flagged articles, report problematic ones via the \href{https://www.irit.fr/~Guillaume.Cabanac/problematic-paper-screener}{PPS}, and request publishers to correct or retract them.

%----------------------------------------------------------------------------------------------------
\subsection{Organisation of an Information Retrieval challenge}
	My background in computing is in Information Retrieval (IR), which has a long experience of experimentation \citep{Voorhees2007}.
	Experimentation in IR requires a coordinating institution (e.g., NIST in the US) to specify a challenging `search task' and invite research groups worldwide to contribute and crack the case.
	The algorithms produced by each participant get benchmarked against a test collection set up by organisers.
	The IR field advances by assessing the relative performance of the proposed methods.

	The collection of problematic papers we grow daily is a valuable by-product of our error detection endeavour.
	I plan to set up an error-detection task at the next iteration of the \emph{Bibliometric-enhanced Information Retrieval} I co-organised the past 7 years \citep{FrommholzEtAl2021c}.
	This will focus the attention of leading IR groups on this research integrity issue and their contributions will help to flag new misconducts.
	Such challenges are meeting opportunities for academics and scientists from the private sector.
	Elsevier is willing to support us and Philippe Terheggen, Managing Director, liaised with us shortly after Elsevier released a public statement in \emph{Retraction Watch} \citep{Marcus2021a}.

%----------------------------------------------------------------------------------------------------
\subsection{Coordination of a distributed `screen and report' task force}
	Our study was framed as an Open Call for Investigation to welcome the community to join forces with us \citep{CabanacEtAl2021c}.
	Now we are leading a crowdsourced effort to (re)assess via the \href{https://www.irit.fr/~Guillaume.Cabanac/problematic-paper-screener}{PPS} papers published with errors.
	About 1,400 articles flagged with 3+ tortured phrases are awaiting for visual inspection and commenting on PubPeer when suspected flaws get confirmed.
	This number increases as we add more tortured phrases to the screener, snowballing from PubPeer.
	We need to co-ordinate this (re)assessment effort and the way Wikipedia moderation works is inspiring.
	Area editors could be appointed to process the queue matching their expertise, and delegate to trusted parties part of the workload.

	Science is said to be self-correcting.
	For this to happen, dedicated people do need to identify issues and correct the record.
	Authors do it when citing previous studies critically.
	We need to bring the critical post peer-review reports to the attention of the publishers.
	This we did about the SCIgen papers \citep{CabanacAndLabbe2021a} and publishers like ACM, IEEE, and Springer retracted the nonsensical papers after a few months.
	For papers not entirely nonsensical, the investigation took longer and is still hanging for some \citep{ByrneEtAl2021a}.
	We should adopt a more systematic approach to notify publishers according to the \href{https://publicationethics.org/guidance/Guidelines}{COPE guidelines} and monitor each case, reactivating our requests to investigate the cases raised.
	This work should benefit from and contribute to the \href{https://retractionwatch.com/retraction-watch-database-user-guide/}{Retraction Watch database}.

%----------------------------------------------------------------------------------------------------
\subsection{Knowledge and skill transfer to the publishing industry}
	Following the publication of \citep{CabanacAndLabbe2021a} integrity managers at various publishers have approached us.
	They typically ask to train their staff to error sleuthing and ask how to use the open source code we had released.
	We consider that better checks should be implemented at the publishers' side to reduce the stream of questionable papers being published.
	Contracting between universities and publishers is a way to pass on knowledge via master classes or professional workshops.
	Some co-authors gained experience in this area and were successful at integrating their screening software into the publishers' pipelines \citep{WeissgerberEtAl2021a}.
	Conversely, we may learn from the experience of publishers and preprint repositories.
	The honorable \href{https://arxiv.org}{arXiv.org} screens submissions with text mining \citep{Ginsparg2014} and it was found to incorrectly accept \href{https://pubpeer.com/publications/74CABE5666BFFCB6D0F696EC1C540A}{one SCIgen paper} only.
	Tying links with the publishing industry would help to populate the collection of problematic papers required to enhance the screening process too (back to step~1 in Sect.~\ref{sec:algo}).

%----------------------------------------------------------------------------------------------------
\subsection{Foster continuing professional education on research misconduct}
	Throughout the IUF tenure, I plan to follow and take part to the communications and activities of \href{https://publicationethics.org/about/our-organisation}{COPE} (Committee on Publication Ethics) and \href{https://irafpa.org}{IRAFPA} (Institut de Recherche et d'Action sur la Fraude et le Plagiat) \citep{BergadaaAndPeixoto2021}.
	I intend to keep communicating about research integrity and misconduct in professional and lay venues \citep{Cabanac2021a}.
	\citet{BergstromAndWest2020} stress the importance of critical thinking when presented statistical analyses not to be deceived.
	Likewise, I believe scientists should consider research integrity as part of their continuing professional education.
	All researchers should take on the responsibility to decontaminate the scientific literature to pass a more sustainable scientific environment to the next generation of scholars.

%=========================================================================================================================
\section{Possible opening of IUF project towards an ERC project during IUF delegation}
During the COVID19 pandemic in 2020, scientific controversies crossed the academic sphere and spilled throughout the public sphere. The ``Surgisphere scandal'' \citep{Piller2020} represented inflated expectations for public health that relied on an information bubble stemming from a flagship journal: The Lancet. The problematic papers that had published unsupported claims sparked much hope, fuss, and confusion before being retracted eventually.

During the national lockdowns, our group of \href{https://nanobubbles.hypotheses.org/team-members}{9 researchers} in France and the Netherlands worked remotely on the issue of apparent defunct self-correction mechanisms of science. We combined perspectives and skills from our diverse backgrounds in engineering (Computing, Nanobiology, and Physics) and in the social sciences (Science and Technology Studies, Sociology of Science). Our submission to the ERC Synergy call has been granted 8.3 million euros for a 5-year period starting in June 2021. 

In a nutshell, Nanobubbles is concerned with how, when and why science fails to correct itself. We are is looking at ways the scientific record can be corrected. Sometimes scientists make mistakes. The way scientists often think about science and non-scientists too is that those mistakes will eventually be cleared up but in practice, people who try to correct mistakes on the scientific record often experience a lot of resistance. We want to see how claims that need to be corrected circulate through scientific communities and also what happens when people contest those claims and try to get them corrected.

Nanobubbles focuses on 3 already identified bubbles (inflated expectations) in the field of nanobiology (see the executive summary \href{https://cordis.europa.eu/project/id/951393}{online}). These bubbles reflect questionable claims that the physics co-PI will reassess by replicating the experiments in his lab (a 6-month demanding research requiring staff trained in biology and physics and complex instrumentation).

During the 5-year IUF tenure, I plan to prepare an application to the ERC Advanced call. AI-powered text generation is not mature enough yet to produce sound scientific texts. The few attempts that have been discovered so far bear rhetorical errors that readers can identify. With the announced and expected advances \citep{Romero2021} in this area, I believe that the next avalanche of problematic papers will stem from text generation with neural networks trained on the entire scientific corpus released in the public domain a few days ago \citep{Else2021d}. Research will be needed to provide editors and reviewers with software tools that detect suspect passages, checking claims against knowledge bases (like \href{https://blast.ncbi.nlm.nih.gov/}{BLAST} for instance), and fabricated data.

%=========================================================================================================================
\section{Possible opening of IUF project towards a project of innovation in teaching methods and dissemination throughout society}
As a researcher, I design and implement information systems to process scholarly big data using text and data mining. These are built on top of databases I modelled and populated using public open data (e.g., Crossref) and subscription-based data acquired through APIs (application programming interfaces). They host 90+ million bibliographic records described with rich metadata, fulltexts, and citation links. End users browse and analyse these data using online dashboards such as the \href{https://www.irit.fr/~Guillaume.Cabanac/problematic-paper-screener}{Problematic Paper Screener} and the \href{https://www.irit.fr/~Guillaume.Cabanac/covid19-preprint-tracker}{COVID19 Preprint Monitor}.

I approach my research and teaching activities as a continuum. The experience I gain while data harvesting, processing, and visualising myself fertilises the lectures I give and the projects I assign students to. The 9-month project I assigned to my 3rd-year students in big data (LP GTIDM) require to process thousands subtitle files of 128 TV series to build a search and recommender system. This topic appeals to them yet I envision a future project of data and text mining of the scientific literature for them to reproduce the \href{https://www.irit.fr/~Guillaume.Cabanac/problematic-paper-screener}{Problematic Paper Screener}.

Education to error detection in the press and social media is part of educational programmes in universities worldwide. The \href{https://www.callingbullshit.org/syllabus.html}{syllabus} and book ``Calling Bullshit: The art of skepticism in a data-driven world'' by \citet{BergstromAndWest2020} of the University of Washington is remarkable. Such training should be part of doctoral programmes (training available from the \href{http://urfist.univ-toulouse.fr/urfist/reseau-urfist}{URFIST national network} I am involved in as \href{https://sygefor.reseau-urfist.fr/#/trainers/extern}{instructor}), including error / misconduct / fraud detection in scientific manuscripts (when readers act as reviewers) and published papers.

Exposure to selected \href{https://pubpeer.com/}{PubPeer} posts and \href{https://retractionwatch.com/}{Retraction Watch} articles could lead doctoral students, tenured faculty, and concerned citizen scientists to reflect on the self-correcting processes in science. Sometimes these work and peer-review corrects flaws before publication or publishers issue corrections and retractions (e.g., \href{https://retractionwatch.com/retracted-coronavirus-covid-19-papers/}{183 retracted} COVID19 studies). Sometimes publishers and authors long to correct the scientific record, which jeopardises the trust the public places in science.

The unstable and ever-evolving nature of the body of knowledge has been heavily commented and sometimes criticised during the pandemic. The public needs a clearer comprehension of how research works (including error detection) and how science accumulates knowledge. Journalists have been disseminating the first results of my project on error detection to the society through radio broadcasts (e.g., \href{https://www.francetvinfo.fr/replay-radio/le-billet-vert/comment-reperer-les-fausses-etudes-scientifiques_4650227.html}{FranceInfo}) and press articles (e.g., \href{https://www.lemonde.fr/sciences/article/2021/06/16/un-nouvel-outil-pour-traquer-les-articles-scientifiques-bidons-ecrits-par-des-logiciels_6084326_1650684.html}{\emph{Le Monde}}, \href{https://en.wikipedia.org/wiki/Neue_Z%C3%BCrcher_Zeitung}{\emph{Neue Zürcher Zeitung}}, \href{https://www.faz.net/aktuell/karriere-hochschule/hoersaal/franzoesische-forscher-entdecken-243-nonsens-papiere-17602613.html}{\emph{Frankfurter Allgemeine Zeitung}}, \href{https://www.nature.com/articles/d41586-021-02134-0}{\emph{Nature}}). During my 5-year tenure as IUF Chair, I wish to keep raising the awareness on this issue while welcoming citizen scientists to contribute to the crowdsourcing of error detection we have started operating with the \href{https://www.irit.fr/~Guillaume.Cabanac/problematic-paper-screener}{Problematic Paper Screener}.

%===============================================================================
\section{Conclusion}
In a nutshell, this application aims to consolidate a growing expertise on error detection in the scientific literature and develop software tools to filter our and even prevent any deceptive flaws from entering the scholarly record.

%===============================================================================
\bibliographystyle{apacite}
\interlinepenalty=10000 % http://tex.stackexchange.com/a/51259
\bibliography{/Users/cabanac/Documents/Recherche/Bibliographies/scientometrics.bib,/Users/cabanac/Documents/Recherche/Bibliographies/annotations.bib,/Users/cabanac/Documents/Recherche/Bibliographies/ri.bib}
\end{document}